\documentclass[prd,twocolumn]{revtex4}
\usepackage{bm}
\usepackage{amssymb}
\usepackage{graphicx}
\begin{document}
\title{Limitations of the adiabatic approximation to the gravitational
       self-force}  
\author{Adam Pound, Eric Poisson, and Bernhard G.\ Nickel}
\affiliation{Department of Physics, University of Guelph, Guelph,
Ontario, Canada N1G 2W1}
\date{November 28, 2005}  
\begin{abstract}
A small body moving in the field of a much larger black hole and
subjected to its own gravity moves on an accelerated world line in the  
background spacetime of the large black hole. The acceleration is
produced by the body's gravitational self-force, which is constructed 
from the body's retarded gravitational field. The adiabatic
approximation to the gravitational self-force is obtained instead 
from the half-retarded minus half-advanced field. It is much easier to 
compute, and it is known to produce the same dissipative effects as
the true self-force. We argue that the adiabatic approximation is
limited, because it discards important conservative terms which lead
to the secular evolution of some orbital elements. We argue further
that this secular evolution has measurable consequences; in
particular, it affects the phasing of the orbit and the phasing of the 
associated gravitational wave. Our argument rests on a simple toy
model involving a point electric charge moving slowly in the weak
gravitational field of a central mass; the charge is also subjected to
its electromagnetic self-force. In this simple context the true
self-force is known explicitly and it can cleanly be separated into
conservative and radiation-reaction pieces. Its long-term effect on
the particle's orbital elements can be fully determined. In this model
we observe that the conservative part of the self-force produces a
secular regression of the orbit's periapsis. We explain how the
conclusions reached on the basis of the toy model can be extended to
the gravitational self-force, and we attempt to extend them also to
the case of rapid motions and strong fields. While the limitations of
the adiabatic approximation are quite severe in a post-Newtonian
context in which the motion is slow and the gravitational field weak,
they may be less severe for rapid motions and strong fields.            
\end{abstract}
\maketitle

\section{Introduction}  

The gravitational inspiral of a solar-mass compact object into a
massive black hole residing in a galactic center has been identified
as one of the most promising sources of gravitational waves for the
Laser Interferometer Space Antenna \cite{LISA}. The need for accurate  
theoretical models of the expected signal, for the purposes of signal 
detection and source identification, has motivated an intense effort
from many workers to determine the motion of the small body in the
field of the large black hole. This is done in a treatment that goes
beyond the geodesic approximation and takes into account the body's
own gravitational field, which is a small perturbation over the field
of the black hole; this must be done without relying on slow-motion or 
weak-field approximations. In this treatment the small body can be
thought of as moving on a geodesic of a perturbed spacetime, or 
equivalently, it can be thought of as moving on an accelerated world 
line in the background spacetime of the large black hole. The second
point of view is generally adopted, and the body is said to move under
the influence of its own gravitational self-force \cite{mino-etal:97,
quinn-wald:97}. The self-force is derived from the retarded
gravitational perturbation produced by the moving body. For a review
of the self-force formalism, see Ref.~\cite{poisson:04b}.    

The concrete evaluation of the gravitational self-force acting on a
small body moving in the Kerr spacetime is a challenging project that
has not yet been completed. Part of this challenge is concerned with
the reconstruction of the metric perturbation \cite{chrzanowski:75a, 
ori:03, lousto-whiting:02} from the Teukolsky variables, which can be
more practically evaluated \cite{teukolsky:73}. Another is concerned
with the regularization of the body's retarded field near the world
line \cite{detweiler-whiting:03, barack-etal:02, barack-ori:03a,
barack-ori:03b, mino-etal:02}, in a context where it is expressed as
an infinite sum over spherical-harmonic modes. Yet another challenge
resides in the fact that the gravitational self-force is a
gauge-dependent quantity \cite{barack-ori:01}; this implies that the
improved equations of motion must be incorporated in a self-consistent
wave-generation formalism before they can yield gauge-invariant
waveforms.     

\section{Adiabatic approximation\ldots}  

Given this state of affairs, it is tempting to seek approximations  
to the self-force formalism that may bypass some of these challenges 
but still produce acceptably accurate results. One such approximation   
was formulated by Mino \cite{mino:03, mino:05a, mino:05b}, who showed
that the long-term evolution of the three principal orbital elements
(the body's orbital energy, angular momentum, and Carter constant,
which would all be constant under geodesic motion but instead evolve
under self-forced motion) can be reproduced on the basis of a
{\it radiation-reaction force} constructed from the half-retarded 
minus half-advanced solution to the perturbation equations. In
particular, Mino showed that the total work done by the
radiation-reaction force equals the total gravitational-wave energy
radiated by the moving body, a result that was previously established
by Quinn and Wald \cite{quinn-wald:99} and Gal'tsov
\cite{galtsov:82}. Mino was further able to show that these long-term
evolutions are, to a very good approximation, gauge-invariant. 

Because the retarded and advanced fields are equally singular on the
world line, their subtraction produces a regular field, and the
computation of the radiation-reaction force requires no
regularization. This computation is therefore much simpler than the
calculation of the true self-force, and the gauge-invariant nature of
the radiation-reaction force argues in favor of its direct involvement 
in waveform calculations.  

The approximation of the true self-force by the radiation-reaction   
force discards the rapid, oscillatory changes of the three principal
orbital elements which occur on the time scale of the orbital period;
it retains only the slow changes that survive after
time-averaging. This replacement of the true self-force by the
radiation-reaction force in a calculation of the orbital evolution is
called the {\it adiabatic approximation}. The computation of inspirals
in the adiabatic approximation, and of their associated
gravitational-wave signals, has so far been pursued by at least two
research groups \cite{drasco-etal:05, hughes-etal:05,
drasco-hughes:05, sago-etal:05}.      

Confidence in the reliability of the adiabatic approximation rests   
on the results of Mino (which are fully reliable) and the belief that   
{\it all secular effects} associated with the true self-force are  
embodied in the long-term evolution of the three principal orbital 
elements (energy, angular momentum, and Carter constant). Because the   
true self-force differs from the radiation-reaction force by purely
conservative terms (which do no work), this is a statement of belief
that the conservative part of the true self-force produces only
short-term effects that do not accumulate over time and do not survive
after time-averaging. This belief is expressed, for example, by
Drasco, Flanagan, and Hughes \cite{drasco-etal:05}, who state near the 
beginning of their Sec.~1.2: ``The effect of the dissipative pieces
of the self force will accumulate secularly, while the effect of the
conservative pieces will not. Hence the effect of the dissipative
pieces on the phase of the orbit will be larger than that of the
conservative pieces by a factor of the number of cycles of inspiral.'' 

\section{\ldots And its limitations} 

Our purpose in this paper is to show that this belief is not well
founded: The conservative terms in the true self-force do produce
effects that accumulate over time and significantly affect the phasing
of the orbit. Our methods allow us to conclude that these effects are 
important in slow-motion, weak-field situations. They do not, however,
allow us to investigate directly more interesting situations involving
rapid motions and strong fields. Nevertheless, a tentative
extrapolation of our slow-motion results indicate that these effects
may be less important in strong-field situations.     

While the radiation-reaction force captures the secular evolution of
the three principal orbital elements (energy, angular momentum, and
Carter constant), it makes no statement regarding the evolution of the
remaining three orbital elements. (We shall call these, for reasons
that will become clear presently, the {\it positional orbital
elements}.) In particular, the radiation-reaction force does not
capture the secular evolution of the positional elements.  

In the case of geodesic motion all six orbital elements are constants
of the motion. The three principal elements serve to label the
geodesic, and the three positional elements serve to specify the
body's initial position on this geodesic. Under orbital evolution
driven by the true self-force, all six elements vary with time, and
the  motion is tangent to an evolving {\it osculating geodesic}. While
it is clear that the principal elements must evolve secularly, we
intend to show that the positional elements also can evolve secularly,
and that this evolution is driven by the conservative part of the true
self-force. The adiabatic approximation, therefore, does not account
for the secular evolution of the positional elements, and we intend to
show that this leads to measurable phasing effects in the waves.    

\section{Electromagnetic self-force in a weakly curved spacetime}   

Our argument is based on a simple toy problem in which the true 
self-force can be evaluated explicitly and decomposed cleanly into 
conservative and radiation-reaction pieces. The toy problem involves a
point electric charge $q$ of mass $m$ moving slowly in the weak
gravitational field of a (noncompact) star of mass $M$. In the toy
problem the charge is a substitute for the small compact body, the
electromagnetic radiation is a substitute for the gravitational
radiation, and the star is a substitute for the massive black hole. We
take the star to be immobile and we place it at the origin of a
three-dimensional Cartesian coordinate system $\bm{x} = [x,y,z]$. The
electromagnetic self-force acting on the point charge was calculated
by DeWitt and DeWitt \cite{dewitt-dewitt:64}, building on the
foundations laid by DeWitt and Brehme \cite{dewitt-brehme:60}; this
calculation was reviewed recently in
Ref.~\cite{pfenning-poisson:02}. Adopting a three-dimensional, 
Newtonian language throughout this section, the equations of motion
are    
\begin{equation} 
m \frac{d^2 \bm{x}}{dt^2} = m \bm{g} + \bm{f}_{\rm self},  
\label{1}
\end{equation} 
where $\bm{x}(t)$ is the charge's position vector, 
\begin{equation} 
\bm{g} = -\frac{M}{r^2} \bm{\hat{r}} 
\label{2}
\end{equation} 
is the Newtonian gravity of the central mass (with $r = |\bm{x}|$ and 
$\bm{\hat{r}} = \bm{x}/r$), and  
\begin{equation} 
\bm{f}_{\rm self} = \lambda_{\rm c} \frac{q^2 M}{r^3} \bm{\hat{r}}  
+ \lambda_{\rm rr} \frac{2}{3} q^2 \frac{d \bm{g}}{dt} 
\qquad (\lambda_{\rm c} = \lambda_{\rm rr} = 1)  
\label{3}
\end{equation} 
is the electromagnetic self-force. The first term on the right-hand  
side of Eq.~(\ref{3}) is the conservative part of the self-force;
the second term is the usual expression for the electromagnetic
radiation-reaction force. We work with units such that $G = c = 1$,
and we have inserted parameters $\lambda_{\rm c} \equiv 1$, 
$\lambda_{\rm rr} \equiv 1$ in order to later distinguish between 
conservative and radiation-reaction effects. In Appendix A we show 
that Mino's prescription \cite{mino:03} for the radiation-reaction
force gives rise precisely to the second term of Eq.~(\ref{3}). The
adiabatic approximation to the electromagnetic self-force therefore
consists of setting $\lambda_{\rm c} = 0$ and $\lambda_{\rm rr} = 1$
in Eq.~(\ref{3}).     

Under the action of $\bm{g}$ alone the point charge would trace a  
Keplerian orbit of semilatus rectum $p$ and eccentricity $e$, and its
orbital radius would be described by 
\begin{equation} 
r = \frac{p}{1 + e \cos(\phi - \omega)}, 
\label{4}
\end{equation}
where $\phi$ is the charge's true longitude and the constant $\omega$
is the longitude at periapsis. In Eq.~(\ref{4}) only $r$ and $\phi$
depend on time. The charge's energy per unit mass is 
$E = -M(1-e^2)/(2p)$ and its angular momentum per unit mass is 
$L = \sqrt{Mp}$. The orbital elements $(p,e)$ therefore act as a 
substitute for $(E,L)$ and can be adopted as the principal orbital
elements. The remaining element $\omega$ is a positional orbital
element. [A second positional element, $t_0$, does not appear
explicitly in Eq.~(\ref{4}) and will not be needed in the sequel; it
is defined by the statement $\phi(t_0) = \omega$.] Because the
self-force of Eq.~(\ref{3}) keeps the orbital motion in a fixed plane,
there is no need to introduce additional orbital elements.    

Under the perturbation produced by the self-force the orbital motion
is still described by Eq.~(\ref{4}) but the elements $(p,e,\omega)$
acquire a time dependence; the motion is tangent to an evolving  
osculating Keplerian orbit. Employing standard methods from celestial  
mechanics we show in Appendix B that the long-term evolution of the 
orbital parameters is governed by the differential equations 
\begin{eqnarray} 
\langle \dot{p} \rangle &=& -\lambda_{\rm rr} 
\frac{4}{3} \frac{q^2 M}{m} \frac{(1-e^2)^{3/2}}{p^2}, 
\label{5} \\ 
\langle \dot{e} \rangle &=& -\lambda_{\rm rr} 
\frac{q^2 M}{m} \frac{e(1-e^2)^{3/2}}{p^3}, 
\label{6} \\ 
\langle \dot{\omega} \rangle &=& -\lambda_{\rm c} 
\frac{1}{2} \frac{q^2 M}{m} \sqrt{\frac{p}{M}} 
\frac{(1-e^2)^{3/2}}{p^3}.   
\label{7}
\end{eqnarray} 
Here the overdot indicates a time derivative, and the angular brackets 
mean that $(\dot{p}, \dot{e}, \dot{\omega})$ are averaged over a
complete period of the unperturbed orbit. It is assumed that 
$q^2 \ll m p$ throughout the evolution, so that the time scales  
associated with the changes in $(p,e,\omega)$ are much longer 
than the orbital period.   

\begin{figure}
\includegraphics[angle=-90,scale=0.33]{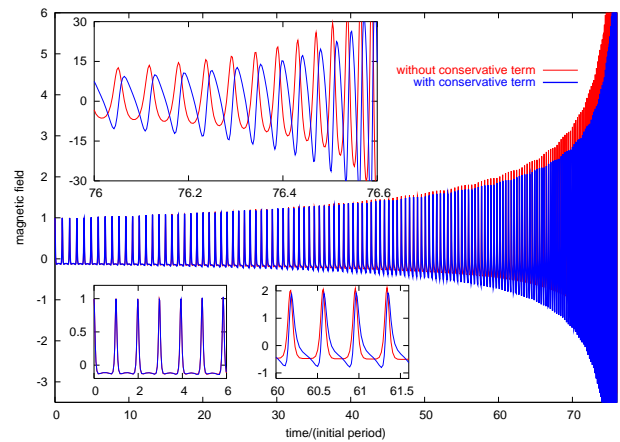}
\caption{Magnetic field produced by the orbiting particle, evaluated
as a function of time in the wave zone, within the electric-dipole
approximation. The units of the magnetic field are arbitrary, and the
time is given in units of the initial orbital period. The initial
orbital elements are $p_0 = 20M$, $e_0 = 0.5$, and 
$\omega_0 = 0$. We set $q^2/(m p_0) = 0.002$. In the color version of 
the figure, the blue curves represent the wave generated by an orbital 
evolution driven by the true self-force, and the red curves represent
the wave calculated within the adiabatic approximation. In the
black-and-white version the true wave appears thick and dark, while
the approximate wave appears thin and light. The main plot shows the
entire inspiral. The fact that the red/light curve appears to move
upward relative to the blue/dark curve is a consequence of a relative
phase shift driven by the secular evolution of $\omega$. The lower
inset on the left shows the earliest part of the inspiral, before the
phase shift had a chance to accumulate. The lower inset on the right
shows a later portion of the inspiral, with the approximated wave
(red/light) leading slightly in phase with respect to the true wave 
(blue/dark). Finally, the upper inset shows the latest portion of the
inspiral, when the two waves are very much out of phase.}   
\end{figure}

Equations (\ref{5})--(\ref{7}) describe the secular evolution of the
orbital elements; they average out the unimportant oscillatory
behavior that leads to a zero cumulative change over a large
number of orbits. The decay of the principal elements $p$ and $e$ 
describes a shrinkage of the orbit accompanied by circularization. 
The presence of $\lambda_{\rm rr} \equiv 1$ in Eqs.~(\ref{5}) and
(\ref{6}) shows that these effects are purely dissipative, and
that the evolution of $p$ and $e$ is driven entirely by the
radiation-reaction part of the self-force. This evolution would be
reproduced exactly by the adiabatic approximation. The negative sign 
on the right-hand side of Eq.~(\ref{7}) indicates that the orbit
undergoes also a regression of its periapsis; the presence of 
$\lambda_{\rm c} \equiv 1$ shows that this effect is purely
conservative. The evolution of $\omega$ is driven entirely by the 
conservative part of the self-force, and this evolution would not be
reproduced by the adiabatic approximation.    

We see that the conservative term in the electromagnetic self-force is
indeed responsible for a secular effect, the accumulating regression
of the orbit's periapsis. This effect would be missed altogether by
the adiabatic approximation, which fails to account for the secular
evolution of the positional orbital elements.  

The secular evolution of $\omega$ has a direct effect upon the phasing 
of the orbit, and upon the phasing of the associated 
electromagnetic-wave signal. This statement is illustrated in Fig.~1, 
which displays the electromagnetic wave generated by the orbiting 
particle; the calculations behind this figure are presented in 
Appendix C. The figure shows very clearly that a model waveform 
calculated on the basis of the adiabatic approximation (which ignores
the evolution of $\omega$) gradually goes out of phase with the true
signal. This is in spite of the fact that both waves incorporate the
same correct long-term evolution of $p$ and $e$.    

The time scale associated with the dephasing can be estimated on the   
basis of Eq.~(\ref{7}). Assuming that $e$ is not too close to unity, 
this is given by $\langle \dot{\omega} \rangle \sim 1/\tau_{\rm dph}$,
or    
\begin{equation} 
\tau_{\rm dph} \sim M \biggl( \frac{mM}{q^2} \biggr) 
\biggl( \frac{p}{M} \biggr)^{5/2}. 
\label{8}
\end{equation} 
The radiation-reaction time scale, on the other hand, can be estimated  
on the basis of Eq.~(\ref{5}). This is given by $\langle \dot{p}
\rangle \sim p/\tau_{\rm rr}$, or 
\begin{equation} 
\tau_{\rm rr} \sim M \biggl( \frac{mM}{q^2} \biggr) 
\biggl( \frac{p}{M} \biggr)^{3}. 
\label{9}
\end{equation} 
The dephasing time is {\it shorter} than the radiation-reaction time 
by a factor of order $\sqrt{M/p} \ll 1$. The total phase shift 
$\Delta \Phi$ accumulated during a radiation-reaction time is
estimated as $\Delta \Phi \sim \langle \dot{\omega} \rangle 
\tau_{\rm rr} \sim \tau_{\rm rr}/\tau_{\rm dph}$, or 
\begin{equation}
\Delta \Phi \sim \sqrt{p/M} \gg 1 
\label{10} 
\end{equation}
after using Eqs.~(\ref{8}) and (\ref{9}). The accumulated phase shift
is large, and this indicates that the adiabatic approach will 
{\it not} produce an adequate approximation of the true waveform over
such a time interval.    

\section{Discussion} 

Our two main conclusions are these:  
\begin{itemize}
\item The secular evolution of the orbital elements is not driven only
  by the radiation-reaction part of the self-force; the conservative
  terms participate also. In particular, the conservative part of the
  self-force gives rise to a secular evolution of the positional
  elements.  
\item The secular evolution of the positional elements, which is not  
  accounted for by an adiabatic approximation to the self-force, has
  measurable consequences. In particular, this evolution affects the
  phasing of the orbit and therefore the phasing of the associated
  wave. 
\end{itemize}
These conclusions apply to the toy problem presented in the preceding
section, but we firmly believe that they are not limited to this
specific example. We are convinced that they hold in the context of
the gravitational self-force acting on a small body moving in the
field of a Kerr black hole. 

Returning to the gravitational problem, we emphasize that it should
not come as a surprise that the conservative terms in the self-force
can drive the secular evolution of the positional orbital
elements. And it should not come as a surprise that this evolution
can produce a measureable effect on the phasing of the gravitational 
waves. We shall elaborate on this observation in the next two
paragraphs, but we note first that a similar point was made recently 
by Ajith {\it et al.}\ \cite{ajith-etal:05a, ajith-etal:05b}.    

Consider the post-Newtonian description of a two-body
system, in a regime where $m$ (the small mass) is much smaller than
$M$ (the large mass). The equations of motion for the relative orbit
take the schematic form   
\begin{equation} 
\bm{a} = \bm{g}_{\rm 0PN} + \bm{g}_{\rm 1PN} + \bm{g}_{\rm 2PN} 
+ \bm{g}_{\rm 2.5PN} + \cdots,   
\label{11}
\end{equation} 
in which $\bm{a}$ is the acceleration of the relative position vector,  
$\bm{g}_{\rm 0PN}$ stands for the Newtonian gravitational field, and
the additional terms are labeled by their post-Newtonian
order. Each term in the post-Newtonian expansion can be expanded in
powers of $m/M$, so that  
\begin{equation} 
\bm{g}_{\rm nPN} = \bm{g}^0_{\rm nPN} 
+ \frac{m}{M} \bm{g}^1_{\rm nPN} + O(m^2). 
\label{12}
\end{equation} 
Summation over post-Newtonian orders of the terms  
$\bm{g}^0_{\rm nPN}$ gives rise to an acceleration $\bm{a}^0$ which 
describes the geodesic motion of a test mass in the gravitational   
field of the mass $M$. Summation of the terms 
$(m/M)\bm{g}^1_{\rm nPN}$ gives rise to $(m/M) \bm{a}^1$, the
correction to the geodesic motion that comes from the true self-force.   
These considerations show that in a post-Newtonian context, the
self-force has conservative terms at order 0PN, 1PN, 2PN, and higher,
and it has dissipative terms at order 2.5PN and higher.   

It is now easy to see why the conservative part of the self-force
must drive a secular evolution of the positional elements. It is
known, for example, that the test-mass term $\bm{g}^0_{\rm 1PN}$
drives a secular shift in $\omega$ relative to Keplerian motion; this
is exemplified by the famous perihelion advance of Mercury. It is also
known that $(m/M)\bm{g}^1_{\rm 1PN}$, the self-force term, contributes
a correction of order $m/M$ to this shift
\cite{damour-deruelle:86}. Concretely, the post-Newtonian expression
for the periapsis advance is given by Eq.~(12.29) of
Ref.~\cite{will:93},  
\begin{equation} 
\langle \dot{\omega} \rangle_{\rm 1PN} =
\frac{3(1-e^2)^{3/2}(M+m)^{3/2}}{p^{5/2}}; 
\label{13} 
\end{equation} 
since this depends on the total mass $M+m$, there is a test-mass 
contribution at order $m^0$ and a self-force contribution at order 
$m$. {\it The self-force therefore produces a secular shift in
$\omega$, over and above the effect already present in the test-mass
limit.} Because the conservative part of the self-force begins at a
low post-Newtonian order, it should be clear that it will drive an
evolution that is potentially more significant than the evolution
driven by dissipative effects, which begin at order 2.5PN. This
property was featured in the toy model: The conservative part of the
electromagnetic self-force can formally be thought of as a 1PN term in
the equations of motion, while the radiation-reaction force is a
smaller 1.5PN term. As we saw this led to a dephasing time that was
shorter (by a factor of the orbital velocity) than the
radiation-reaction time.       

Incidentally, an extension of this argument shows why a recent claim 
made by Burko \cite{burko:04} must be erroneous. Burko claims that if 
the mass $m$ is spinning rapidly, then the effects of the conservative  
spin-orbit interaction will dominate over conservative self-force
effects. It is easy to see why this claim must be false:   
Formally the spin-orbit contribution to Eq.~(\ref{11}) begins
at 1PN order, but for rapidly-spinning compact bodies the
post-Newtonian scaling is in fact shifted to 1.5PN
\cite{kidder-will-wiseman:93, kidder:95}. Thus the spin-orbit
interaction contributes a term $\bm{g}_{\rm 1.5PN}$, and the fact that 
the spin angular momentum scales as the mass squared implies that 
$\bm{g}_{\rm spin-orbit} = (m/M)\bm{g}^1_{\rm 1.5PN}$. This addition 
to the self-force is therefore of 1.5PN order, and it will not
dominate over the 1PN term that produces the periapsis shift. 

These observations generalize to rapid motions and strong 
gravitational fields. In this context the conservative part of the
self-force should still be expected to drive a secular evolution of
the positional orbital elements --- those that characterize the
initial position of the small body on the osculating geodesic. The
secular evolution of the positional elements will affect the phasing
of the orbit and the phasing of the associated gravitational wave. It
is not, however, accounted for by the adiabatic approximation, which 
captures only the dissipative aspects of the orbital evolution. 

We must add that there are important differences between weak-field
and strong-field situations, and admit that our weak-field methods do
not allow us to determine the strong-field effects associated with the 
conservative part of the self-force. We shall, nevertheless, attempt
to draw some conclusions by extrapolating our weak-field results to 
strong-field situations. These conclusions, of course, are tentative
and await confirmation from a proper strong-field investigation.       

Our first observation is that in the case of rapid motions and strong
fields, the conservative-driven evolution occurs over a time scale
that is now comparable with (and not much shorter than) the 
radiation-reaction time scale. The separation of scales no longer
occurs because the post-Newtonian scaling with the orbital velocity
does not apply when the velocity is close to unity. This implies that
the total accumulated phase shift now amounts to a number of wave
cycles that may not be much larger than unity. Indeed, the estimate of
Eq.~(\ref{10}) must be replaced by $\Delta \Phi \sim (p/M)^{3/2}$ in
the case of gravity; this number would be large in a post-Newtonian
situation, but in the present context we have $p/M \agt 1$ and this
leads to $\Delta \Phi \agt 1$. This represents a fractional correction
of order $m/M$ to the total number of wave cycles accumulated during
the entire inspiral, which is of order $M/m$.    

This strong-field estimate of the size of the accumulated phase shift
is in agreement with statements made in Ref.~\cite{hughes-etal:05} and
in Sec.~1.2 of Drasco, Flanagan, and Hughes
\cite{drasco-etal:05}. Recall the statement quoted earlier in Sec.~II:
``The effect of the dissipative pieces of the self force will
accumulate secularly, while the effect of the conservative pieces will
not. Hence the effect of the dissipative pieces on the phase of the
orbit will be larger than that of the conservative pieces by a factor
of the number of cycles of inspiral.'' While we still disagree with
the first sentence, we acknowledge that the second sentence may well
be true in the context of strong fields and rapid motions. But the
fact that $\Delta \Phi$ scales as $(p/M)^{3/2}$ and must therefore
be large in post-Newtonian situations seems to have been overlooked by 
these authors.      

As a final remark we add that our considerations are guided by the   
realistic expectation that the orbital motion of a small compact body
around a massive black hole will be eccentric and taking place outside
the equatorial plane of the rotating black hole. For the special case
of equatorial, circular orbits the relevance of the positional orbital  
elements disappears; the secular drift of $\omega$, for example, has
no measurable consequences for circular orbits. In such special
cases the adiabatic approximation can be expected to be reliable. We
suspect that it is largely the consideration of circular, equatorial
orbits that has boosted the confidence in this approach; refer, for 
example, to Appendix A of Drasco, Flanagan, and Hughes 
\cite{drasco-etal:05}.  
     
\acknowledgments 

This work was supported by the Natural Sciences and Engineering
Research Council of Canada. We have benefited from conversations with
Leor Barack, Steve Detweiler, Bala Iyer, Eran Rosenthal, and Bernard
Whiting. A correspondence with Steve Drasco, Eanna Flanagan, and Scott
Hughes has been essential to the development of the conclusions
presented in this paper.       

\appendix 
\section{Retarded, advanced, and radiation-reaction forces}   

The electromagnetic self-force acting on a point charge $q$ moving in
an arbitrary curved spacetime was first calculated by DeWitt and
Brehme \cite{dewitt-brehme:60}, and their expression was later
corrected by Hobbs \cite{hobbs:68}. We briefly sketch the main steps
of the derivation here, relying heavily on the presentation given in
Sec.~5.2 of Ref.~\cite{poisson:04b} (hereafter referred to as LRR). We
then provide an expression for the advanced version of the
self-force. Subtracting the two gives Mino's radiation-reaction force
\cite{mino:03}. Finally, we evaluate these forces in the case of a
charge moving slowly in the weak gravity of a central mass $M$,
relying heavily on the work of DeWitt and DeWitt
\cite{dewitt-dewitt:64} as reviewed in Ref.~\cite{pfenning-poisson:02} 
(hereafter referred to as PP).   
  
The derivation of the true, retarded, self-force begins with the
following expression for the retarded potential produced by a charged
particle moving on a world line described by the parametric relations
$z^\mu(\tau)$ [LLR Sec.~5.2.2, Eq.~(443)]:
\begin{equation}
A^\alpha_{\rm ret}(x) = q \int
G^{\ \ \alpha}_{{\rm ret}\, \mu} \bigl(x,z(\tau)\bigr) 
u^\mu(\tau)\, d\tau.    
\label{A1}
\end{equation} 
Here $x$ is the spacetime point at which the potential is evaluated, 
$G^{\ \ \alpha}_{{\rm ret}\, \alpha'}(x,x')$ is the retarded Green's
function associated with the wave equation satisfied by $A^\alpha$ (in
the Lorenz gauge), and $u^\mu = dz^\mu/d\tau$ is the charge's velocity
vector.  

The electromagnetic field $F^{\alpha\beta}_{\rm ret}(x)$ obtained from
the retarded potential is singular on the world line, and it must be
regularized before taking the limit $x \to z$ and evaluating the 
Lorentz force $q F^\mu_{\ \nu}(z) u^\nu$. Detweiler and Whiting
\cite{detweiler-whiting:03} have shown that the correct regularization 
procedure is to remove from $F^{\alpha\beta}_{\rm ret}(x)$ a singular
field $F^{\alpha\beta}_{\rm S}(x)$ which is known to exert no force
on the particle. The singular field is defined and evaluated in
Sec.~5.2.5 of LRR, and the regular remainder 
\begin{equation} 
F^{\alpha\beta}_{\rm R}(x) = F^{\alpha\beta}_{\rm ret}(x) 
- F^{\alpha\beta}_{\rm S}(x)
\label{A2}
\end{equation}
is what must be substituted into the Lorentz-force equation. 

When the charged particle moves freely (being subjected to no other
force but its own self-force) in a vacuum region of the spacetime, the
resulting equations of motion are $m a^\mu = f^\mu_{\rm ret}$, where
$m$ is the particle's mass, $a^\mu$ its covariant acceleration, and  
[LRR Sec.~2.5.6, Eq.~(481)] 
\begin{equation} 
f^\mu_{\rm ret}(\tau) = 2 q^2 u_\nu \int_{-\infty}^{\tau^-}      
\nabla^{[\mu} G^{\ \ \nu]}_{{\rm ret}\,\lambda'}
\bigl(z(\tau),z(\tau')\bigr) u^{\lambda'}(\tau')\, d\tau'
\label{A3}
\end{equation}
is the true, retarded, electromagnetic self-force. Notice that the
integration extends over the particle's past history, and that it is
cut short at $\tau' = \tau^- = \tau - 0^+$ to avoid the singular
behavior of the retarded Green's function at coincidence. This
limiting procedure was first derived by DeWitt and Brehme 
\cite{dewitt-brehme:60}, and it is a natural byproduct of the
Detweiler-Whiting regularization method \cite{detweiler-whiting:03}.  

The advanced version of the self-force is obtained by starting 
with the advanced solution for the potential,  
\begin{equation}
A^\alpha_{\rm adv}(x) = q \int  
G^{\ \ \ \alpha}_{{\rm adv}\, \mu} \bigl(x,z(\tau)\bigr) 
u^\mu(\tau)\, d\tau,
\label{A4}
\end{equation} 
and going through the same calculational steps as described above. 
The advanced electromagnetic field $F^{\alpha\beta}_{\rm adv}(x)$ also
is singular on the world line, but it is regularized by the same
singular field $F^{\alpha\beta}_{\rm S}(x)$ as the retarded field. The
end result for the equations of motion is $m a^\mu = f^\mu_{\rm adv}$, 
with 
\begin{equation} 
f^\mu_{\rm adv}(\tau) = 2 q^2 u_\nu \int_{\tau^+}^{\infty}       
\nabla^{[\mu} G^{\ \ \nu]}_{{\rm adv}\,\lambda'}
\bigl(z(\tau),z(\tau')\bigr) u^{\lambda'}(\tau')\, d\tau'
\label{A5}
\end{equation}
being the advanced version of the electromagnetic self-force. Notice
that the integration now extends over the particle's future history,
and that it is cut short at $\tau' = \tau^+ = \tau + 0^+$ to avoid the
singular behavior of the advanced Green's function at coincidence. 

Mino's radiation-reaction force is defined by \cite{mino:03} 
\begin{equation} 
f^\mu_{\rm rr} = \frac{1}{2} \Bigl( f^\mu_{\rm ret} 
- f^\mu_{\rm adv} \Bigr).
\label{A6}
\end{equation}
This depends on the particle's entire history from $\tau' = -\infty$
to $\tau' = +\infty$. Because the retarded and advanced fields are
equally singular on the world line, the evaluation of the
radiation-reaction force does not require regularization. As was
mentioned in Sec.~II of the main text, the total work done by the 
radiation-reaction force equals the total energy radiated by the 
charged particle \cite{quinn-wald:99}.  

The expressions (\ref{A3}), (\ref{A5}), and (\ref{A6}) for the
retarded, advanced, and radiation-reaction forces are valid for any
charge that moves freely in a vacuum region of an arbitrary
spacetime. Let us now specialize to the case of a particle moving
slowly in the weakly curved spacetime of a (noncompact) star of mass
$M$. The steps required to compute the retarded Green's function for
such a spacetime, and to evaluate the integral of Eq.~(\ref{A3}), are 
detailed in PP \cite{pfenning-poisson:02}. They lead to 
[PP Eq.~(5.23)]      
\begin{equation} 
\bm{f}_{\rm ret} = \frac{q^2 M}{r^3} \bm{\hat{r}}  
+ \frac{2}{3} q^2 \frac{d \bm{g}}{dt},  
\label{A7}
\end{equation} 
which is also Eq.~(\ref{3}) of Sec.~IV, where the notation is
introduced. Going through the same steps, starting instead with 
Eq.~(\ref{A5}), gives  
\begin{equation} 
\bm{f}_{\rm adv} = \frac{q^2 M}{r^3} \bm{\hat{r}}  
- \frac{2}{3} q^2 \frac{d \bm{g}}{dt}. 
\label{A8}
\end{equation} 
Not surprisingly, the advanced self-force is obtained from the
retarded self-force by time reversal, $t \to -t$. Finally, from
Eqs.~(\ref{A6})--(\ref{A8}) we obtain Mino's radiation-reaction force,   
\begin{equation} 
\bm{f}_{\rm rr} = \frac{2}{3} q^2 \frac{d \bm{g}}{dt}. 
\label{A9}
\end{equation} 
This is the familiar result that follows from the
Abrahams-Lorentz-Dirac equation in flat spacetime (see, for example,
Ch.~16 of Ref.~\cite{jackson:99}).   

\section{Orbital evolution under the self-force} 

In this Appendix we provide a derivation of
Eqs.~(\ref{5})--(\ref{7}). We rely on standard analytical methods
of celestial mechanics as described, for example, in
Ref.~\cite{danby:88}.   

We work in the $x$-$y$ plane and introduce the polar coordinates
$(r,\phi)$ defined by $x = r\cos\phi$ and $y = r\sin\phi$. We use 
the basis $(\bm{\hat{r}},\bm{\hat{\phi}})$, in which $\bm{\hat{r}}$ is 
a unit vector pointing in the direction of increasing $r$, and
$\bm{\hat{\phi}}$ is a unit vector pointing in the direction of
increasing $\phi$. The particle's velocity vector, for example, can be 
decomposed as $\bm{v} = \dot{r} \bm{\hat{r}} + (r\dot{\phi})
\bm{\hat{\phi}}$, where a dot indicates differentiation with respect
to $t$.  

The self-force of Eq.~(\ref{3}) admits the decomposition  
\begin{equation} 
\bm{f}_{\rm self} = m \bigl(R \bm{\hat{r}} + S \bm{\hat{\phi}} \bigr),  
\label{B1}
\end{equation}
with 
\begin{equation} 
R = \frac{q^2 M}{m} \biggl( \lambda_{\rm c} \frac{1}{r^3} 
+ \lambda_{\rm rr} \frac{4}{3} \frac{\dot{r}}{r^3} \biggr) 
\label{B2}
\end{equation} 
and 
\begin{equation} 
S = \frac{q^2 M}{m} \biggl( -\lambda_{\rm rr}
\frac{2}{3} \frac{\dot{\phi}}{r^2} \biggr).  
\label{B3}
\end{equation} 
We treat $q^2/m$ as a small parameter and the self-force as a small  
perturbing force that produces only a slight deviation with respect to 
Keplerian motion. 

The charge's Keplerian orbit is described by Eq.~(\ref{4}), which we
rewrite as 
\begin{equation} 
r = \frac{p}{1 + e \cos v}, 
\label{B4}
\end{equation}
in terms of the true anomaly $v$; the longitude is then given by 
$\phi = v + \omega$. For Keplerian motion the orbital elements
$(p,e,\omega)$ are constant, and $t(v)$ is determined by integrating
the differential equation 
\begin{equation} 
\dot{v} = \sqrt{\frac{M}{p^3}} \bigl(1 + e \cos v \bigr)^2. 
\label{B5} 
\end{equation} 
Combining Eqs.~(\ref{B4}) and (\ref{B5}) produces 
\begin{equation} 
\dot{r} = e \sqrt{\frac{M}{p}} \sin v. 
\label{B6}
\end{equation} 
The orbital period is the time required by the particle to go from 
periapsis ($v=0$) to apoapsis ($v=\pi$) and then back to periapsis 
($v=2\pi$). It is given by 
\begin{equation}
P = 2\pi \sqrt{ \frac{p^3}{(1-e^2)^3 M} }.
\label{B7}
\end{equation} 
During this interval $\phi$ increases from $\omega$ to $\omega +
2\pi$. 

The electromagnetic self-force perturbs the particle's motion away
from its Keplerian orbit. In the method of {\it osculating orbital 
elements} the perturbed motion is still described exactly by
Eqs.~(\ref{B4})--(\ref{B6}), but the elements $(p, e, \omega)$
acquire a time dependence. The driving equations are (see, for
example, Ch.~11 of Ref.~\cite{danby:88})  
\begin{eqnarray} 
\dot{p} &=& 2 \sqrt{\frac{p^3}{M}} \frac{S}{1 + e\cos v},  
\label{B8} \\ 
\dot{e} &=& \sqrt{\frac{p}{M}} \biggl[ R \sin v + S 
\frac{2\cos v + e(1 + \cos^2 v)}{1 + e\cos v} \biggr], 
\label{B9} \\ 
\dot{\omega} &=& \sqrt{\frac{p}{M}} \frac{1}{e} \biggl[ -R \cos v + S
\frac{(2+e\cos v)\sin v}{1 + e \cos v} \biggr],  
\label{B10}
\end{eqnarray}   
where $R$ and $S$ are listed in Eqs.~(\ref{B2}) and (\ref{B3}),
respectively. 

For the purpose of obtaining the {\it secular behavior} of the
orbital elements, it is sufficient to integrate
Eqs.~(\ref{B8})--(\ref{B10}) approximately, employing the method of 
averaging (see, for example, Ch.~12 of Ref.~\cite{kahn:90}). In this
method we average the right-hand side of each equation over a complete
orbital period, treating $(p,e,\omega)$ as constants within the
integral. If, for example, $\dot{I}(p,e,\omega;v)$ denotes the time
derivative of orbital element $I$, then we calculate 
\begin{equation} 
\langle \dot{I} \rangle \equiv \frac{1}{P} \int_0^P 
\dot{I}(p,e,\omega;v)\, dt = \frac{1}{P} \int_0^{2\pi} 
\frac{\dot{I}(p,e,\omega;v)}{\dot{v}}\, dv, 
\label{B11}
\end{equation} 
using Eq.~(\ref{B5}) for $\dot{v}$ and keeping $(p,e,\omega)$ fixed
while integrating. The result discards the rapid oscillations in
$I(t)$ which do not accumulate over a large number of orbits. It
keeps, however, the secular drift that eventually produces a large   
cumulative change. 

Substitution of Eqs.~(\ref{B8})--(\ref{B10}) into the integral of
Eq.~(\ref{B11}) leads to the results displayed in
Eqs.~(\ref{5})--(\ref{7}). We recall that only the radiation-reaction 
component of the self-force is responsible for the secular evolution
of $p$ and $e$, while only the conservative component is responsible
for the secular change in $\omega$.  

We note that in the case of a purely conservative self-force [obtained
by setting $\lambda_{\rm rr} = 0$ in Eq.~(\ref{3})] the relation
between $\phi$ and $v$ can be obtained exactly: $\phi(v) =
[1-q^2/(mp)]^{1/2} v$. This exact expression leads to a modified
form of Eq.~(\ref{7}),  
\begin{equation} 
\langle \dot{\omega} \rangle = -\frac{1}{2} \frac{q^2 M}{m}
\sqrt{\frac{p}{M}} \frac{(1-e^2)^{3/2}}{p^3} f(x),    
\label{B12}
\end{equation}
where $x \equiv q^2/(mp)$ and 
\begin{equation}
f(x) = \frac{1 - \sqrt{1-x}}{x/2} = 1 + \frac{x}{4} + \frac{x^2}{8} +
\cdots. 
\label{B13}
\end{equation} 
When $q^2 \ll mp$ we have that $f(x) \simeq 1$ and Eq.~(\ref{7}) is a
very good approximation to Eq.~(\ref{B12}). 

In Appendix C we will need explicit expressions for $\langle p
\rangle(v)$, $\langle e \rangle(v)$, and $\langle \omega
\rangle(v)$. To obtain these we first average $I' \equiv dI/dv = 
\dot{I}/\dot{v}$ over a complete orbital period, to get $\langle I'
\rangle \equiv (2\pi)^{-1} \int_0^{2\pi} I'\, dv = (P/2\pi) \langle
\dot{I} \rangle$. This yields 
\begin{eqnarray} 
\langle p' \rangle &=& -\lambda_{\rm rr} \frac{4}{3} \frac{q^2}{m}
\sqrt{\frac{M}{p}}, 
\label{B14} \\ 
\langle e' \rangle &=& -\lambda_{\rm rr} \frac{q^2}{m} e
\sqrt{\frac{M}{p^3}}, 
\label{B15} \\ 
\langle \omega' \rangle &=& -\lambda_{\rm c} \frac{1}{2} \frac{q^2}{m} 
\frac{1}{p}.  
\label{B16}
\end{eqnarray} 
Next we integrate, and obtain 
\begin{eqnarray} 
\langle p \rangle &=& p_0 
\bigl(1-\lambda_{\rm rr} v/v_0 \bigr)^{2/3}, 
\label{B17} \\ 
\langle e \rangle &=& e_0
\bigl(1-\lambda_{\rm rr} v/v_0 \bigr)^{1/2}, 
\label{B18} \\ 
\langle \omega \rangle &=& -\frac{3}{4} \sqrt{\frac{p_0}{M}} 
\Bigl[ 1 - \bigl(1-\lambda_{\rm c} v/v_0 \bigr)^{1/3} \Bigr],   
\label{B19} 
\end{eqnarray} 
where 
\begin{equation} 
v_0 = \frac{m}{2q^2} \sqrt{\frac{{p_0}^3}{M}}. 
\label{B20}
\end{equation} 
We assume that the orbital evolution starts at $v=0$ with values
$(p=p_0,e=e_0,\omega=0)$ for the orbital elements. The evolution ends
at $v=v_0$ when both $\langle p \rangle$ and $\langle e \rangle$ are
zero; the final value of the periapsis shift is $\langle \omega
\rangle(v_0) = -(3/4)\sqrt{p_0/M}$, and this expression confirms the 
estimate of Eq.~(\ref{10}).     
  
\section{Electromagnetic wave} 

We calculate the electromagnetic wave (or more precisely, the
wave-zone magnetic field) produced by a charged particle subjected to
the electromagnetic self-force of Eq.~(\ref{3}). As was explained in
Appendix B, the motion is at all times described by $r(v)$ given by
Eq.~(\ref{B4}), $\phi(v) = v + \omega$, and $t(v)$ is obtained by
numerically integrating Eq.~(\ref{B5}). The orbital elements
$(p,e,\omega)$ evolve according to Eqs.~(\ref{B17})--(\ref{B20}).  

In the electric-dipole approximation the wave-zone magnetic field is
given by (see, for example, Sec.~11.1.4 of Ref.~\cite{griffiths:99})   
\begin{equation} 
\bm{B} = -\frac{\mu_0}{4\pi} \frac{\bm{\hat{n}} \times \ddot{\bm{p}}}
{c d}, 
\label{C1}
\end{equation}
where $\bm{\hat{n}} =
[\sin\alpha\cos\beta,\sin\alpha\sin\beta,\cos\alpha]$ is a unit vector  
which points from the source to the detector, and $\bm{p}$ is the
source's electric dipole moment expressed in terms of retarded time
$t-d/c$; the detector is at a distance $d$ from the source and $c$ is
the speed of light. For an orbiting charge we have $\ddot{\bm{p}} = q
\bm{a}$, where $\bm{a}$ is the particle's acceleration. For our
purposes it is sufficient to use the approximation $\bm{a} = \bm{g} =
-(M/r^2) \bm{\hat{r}}$ and Eq.~(\ref{C1}) becomes  
\begin{equation} 
\bm{B} = \frac{\mu_0}{4\pi} \frac{qM}{c d} 
\frac{\bm{\hat{n}} \times \bm{\hat{r}}}{r^2}. 
\label{C2}
\end{equation}
With $\bm{\hat{r}} = [\cos\phi,\sin\phi,0]$ we have 
\begin{equation} 
\bm{\hat{n}} \times \bm{\hat{r}} = [-\cos\alpha \sin\phi, \cos\alpha
  \cos\phi, \sin\alpha \sin(\phi-\beta)],  
\label{C3}
\end{equation}     
and each component of the magnetic field can be computed
straightforwardly.  

To produce the plots of Fig.~1 we chose the viewing angles $(\alpha =
\pi/4, \beta = 0)$ and selected the $y$ component of the magnetic
field. To integrate the equations of motion we set $p_0 = 20M$, $e_0 = 
0.5$, and $q^2/(mp_0) = 0.002$. The magnetic field was rescaled by an
arbitrary numerical factor to obtain values of order unity. To produce
the true wave (represented as a blue/dark curve) we set 
$\lambda_{\rm rr} = \lambda_{\rm c} = 1$ in
Eqs.~(\ref{B17})--(\ref{B19}). To produce the adiabatic approximation
to the wave (represented as a red/light curve) we selected 
$\lambda_{\rm rr} = 1$ and $\lambda_{\rm c} = 0$.  

\bibliography{../bib/master} 
\end{document}